# A dynamical model of remote-control model cars

Álvaro Suárez[1], Daniel Baccino[2] and Arturo C Martí[3]

[1] Departamento de Física, Consejo de Formación en Educación, Montevideo, Uruguay
[2] Departamento de Física, Consejo de Formación en Educación, Montevideo, Uruguay
[3] Instituto de Física, Facultad de Ciencias, Universidad de la República, Montevideo, Uruguay

E-mail: alsua@outlook.com

## Abstract

Simple experiments for which differential equations cannot be solved analytically can be addressed using an effective model that satisfactorily reproduces the experimental data. In this work, the one-dimensional kinematics of a remote-control model (toy) car was studied experimentally and its dynamical equation modelled. In the experiment, maximum power was applied to the car, initially at rest, until it reached its terminal velocity. Digital video recording was used to obtain the relevant kinematic variables that enabled to plot trajectories in the phase space. A dynamical equation of motion was proposed in which the overall frictional force was modelled as an effective force proportional to the velocity raised to the power of a real number. Since such an equation could not be solved analytically, a dynamical model was developed and the system parameters were calculated by non-linear fitting. Finally, the resulting values were substituted in the motion equation and the numerical results thus obtained were compared with the experimental data, corroborating the accuracy of the model.

## 1. Introduction

The kinematic and dynamic aspects associated with the motion of remote-control model (or toy) cars and the electromagnetic aspects associated with the operation and the efficiency of their small built-in electric motor as well as with the transmission and reception of electromagnetic waves for controlling their motion deserve the attention of researchers in and teachers of Physics.

Wick and Ramsdell [1, 2] modelled the motion of toy cars rolling down an arbitrarily defined track. In their experiment, turning points were expressed in terms of height loss relative to a hypothetical frictionless situation based on the static friction coefficient between the car and the track. The authors provided a detailed analysis of different track shapes and the effects of air friction, but failed to account for the effects of rolling. In addition, unlike the remote control cars of our study, the cars used by Wick and Ramsdell were not driven by a built-in motor but rolled by the effect of gravity. In a later work, Wick and Ramsdell [3] studied the motion of an electric toy train. The analysis focused on aspects of friction, electrically induced torque and electromotive forces, and included other effective parameters needed to develop a model that could be solved numerically. Unlike the case with remote-control cars, the power input of a train can be changed arbitrarily by accurately adjusting the voltage delivered by an external regulated source of direct current.

Care must be taken by Physics teachers to avoid misleading students into thinking that the behaviour of real systems can always be described from a purely theoretical perspective and expressed in terms of simple equations that can be solved analytically. As an example, experiments with remote-control cars can be carried out easily but cannot be easily modelled. The concepts and tools necessary for developing a suitable model are described in this paper.

The motion of a remote-control car can be modelled by describing trajectories in the phase space, with the acceleration and the velocity as variables [4]. The kinematic variables can be obtained from the analysis of digital video recording of the car in motion. In this work, use was made of the Tracker video analysis and modelling tool [5, 6] capable of determining the position of a moving object as a function of time and then using it in numerical derivation schemes in order to obtain other magnitudes, such as its velocity or acceleration. Tracker was also used to develop a dynamical model describing the kinematic behaviour of the car. Non-linear fitting to the trajectory in the phase space was used to determine approximate values of the parameters in the motion equation. Tracker also carried out numerical integration of the motion equation, the results of which were plotted and compared with the experimental plots.

The experimental analysis of the evolution of different physical phenomena based on digital video recordings has received particular attention in the literature in the past years [7, 8, 9, 10, 11, among others]. In contrast, the same does not hold true for the use of dynamical models to verify the predictive power of motion equations, the works by Wee [12, 13] being notable exceptions. Clearly, where motion equations can be solved analytically—as is usually the case with laboratory experiments—the numerical solution of models appears to lack didactic value. However, real systems can seldom be modelled from a purely theoretical perspective.

The dynamic behaviour of remote-control toy cars in the phase space cannot be described in terms of equations that can be solved analytically. In this paper, a model based on a non-linear motion equation was found to reproduce the system's behaviour with a satisfactory degree of accuracy. The model was characterized based on the car's trajectories in the phase space in terms of velocity and acceleration, created by Tracker. The predictive power of the model was finally verified by comparing the solution to the differential equation with the car's position at different times.

## 2. Experimental setup

The remote-control toy car was 17 cm in length, 7.3 cm in width, 3.8 cm in height, and 0.1736 kg in mass (Figure 1). The car was capable of moving along a straight line on an even, level surface.

In the experiment, applying maximum power by pulling the remote-control lever, the car started from rest and accelerated until it reached its terminal velocity. The car in motion was filmed with a Kodak PlaySport video camera mounted on a tripod. In order to obtain the sharpest image possible, light spots were used to improve the lighting conditions and reduce the shutter time of the camera. The recording was analysed using the Tracker video analysis and modelling tool. In order to determine the car's position as a function of time, the *autotracker* tool was enabled. This tool is capable of selecting a pattern within a frame and track it across the rest of the recording. Figure 2 shows a screenshot

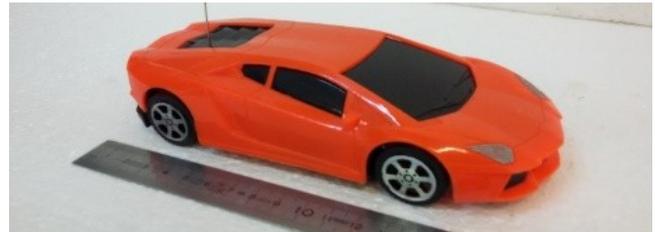

of the *autotracker* interface in use.

Figure 1. Remote-control model car used in the experiments.

Figure 2 *Tracker* screenshot showing car's position, previous marks (red romboids), axes (purple) and calibration tape (blue).

Figure 3. Temporal evolution of the car's position obtained with *Tracker*.

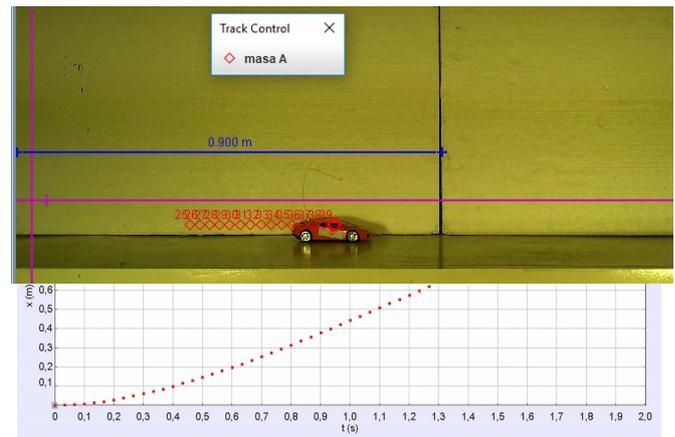

## 3. Experimental results and analysis of kinematic variables

The temporal evolution of the car's position was created using the *autotracker* tool of Tracker. An example is shown in Figure 3. The velocity and acceleration curves shown in Figures 4 and 5 were obtained by numerical derivation performed by Tracker. These curves clearly show that the car approached its terminal velocity asymptotically, consistent with the expected behaviour. The position curve is noticeably smooth, the velocity curve is slightly noisy, and the acceleration curve is markedly noisier.

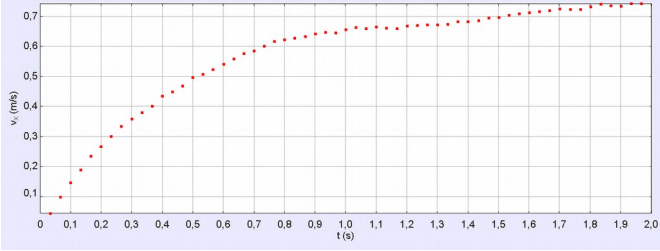

Figure 4. Velocity of the model car as a function of time.

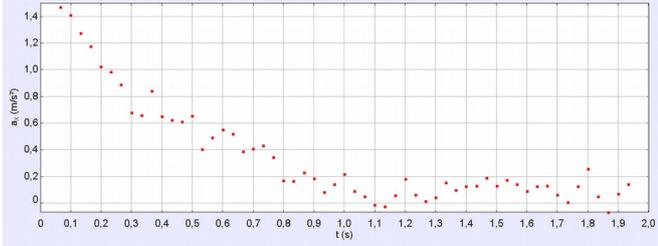

Figure 5. Acceleration of the model car as a function of time.

In the experiments, the car, starting from rest, accelerated at maximum motor power until it reached its terminal velocity. In the direction of motion, the car was accelerated by the frictional force exerted on the car by the surface, and was decelerated by resisting forces due to air friction. The order of magnitude of the forces acting on the car can be estimated from the analysis of the velocity and acceleration curves as a function of time. Irrespective of the model used to describe these forces, the system undergoes a transient state in which the acting forces change with the car's velocity, so that the car, initially at rest, eventually reaches a terminal velocity. At the initial time, as the velocity-dependent frictional force was zero, the net force acting on the car was equal to the net driving force. Based on the initial acceleration and mass of the car, the net driving force was estimated to be of the order of 1.4 m/s$^2$. 0.17 kg ~ 0.2 N. This value is similar to that of the effective dissipative force determined when the car reached its terminal velocity.

Based on the estimation of the effective dissipative force in the stationary state, it is also possible to estimate the relative contribution of the drag forces acting on the car with respect to other sources of dissipation. The air friction force is a complex function even for objects with very simple geometries like spheres or cylinders. Dimensional analysis suggests that the frictional force acting on an object of a given geometry, expressed as a function of the drag coefficient, can be related to the average velocity according to

$$F_d = \frac{1}{2} \rho v^2 C_d A \quad (1)$$

where $\rho$ is the air density, A is the car's frontal surface area and $C_d$ is the drag coefficient [14], which in the case of sports cars is about 0.3 [15].

As the car's terminal velocity was approximately 1.2 m/s, the maximum drag force, based on the car's dimensions, was of the order of 10$^{-3}$ N, amounting to less than 1% of the maximum effective dissipative force. Based on these calculations, it was demonstrated that the effective dissipative force acting on a remote-control toy car originates mainly in its internal mechanisms.

## 4. System dynamics

### 4.1 Dynamical model in the phase space

In order to determine the motion equation for the car. One possible approach would be to characterize the car's motor and to model the dissipative effects associated with internal friction forces within the car. Because of the numerous details that would need to be taken into account, this approach would be excessively time consuming and could not be addressed in basic university settings. Stemming from the notion of phase space, a simpler alternative relies on, the determination of an effective dynamical equation that reproduces the main characteristics of the car's behaviour. In order to develop the model, the force exerted by the surface was represented as the resultant of two forces: one associated with the motor drive, being constant in magnitude in the direction of motion, and the other acting in the opposite direction, being dependent on the velocity and encompassing all the dissipative effects associated with the motor and internal friction. Therefore, the motion equation for the car can be written as

$$M \frac{dv}{dt} = F - k v^n \quad (2)$$

where $M$ is the car's mass, $F$ is the driving force, assumed to be constant, and parameters $k$ and $n$ are real numbers that define the functional dependence of the dissipative force on the velocity. Equation (2) shows that, as the dissipative force increases with the velocity, accelerating the car will eventually lead to a situation of dynamic equilibrium. The car's terminal velocity is given by

$$v_{limit} = \sqrt[n]{\frac{F}{k}} \quad (3)$$

In order to characterize the dynamic behaviour, it is necessary to determine $F$, $k$ and $n$, assuming the mass of the car is known. However, Eq. (2) can be solved analytically only when $n=1$ or $n=2$.

The system dynamics was modelled based on the phase-space trajectory in terms of velocity and acceleration, created by Tracker. Assuming that the car's motion can be suitably described by Eq. (2), the phase space curve was fitted to the following function

$$a = C - D v^n \quad (4)$$

Figure 6 shows the fit of the data recorded in the *a(v)* phase space. Knowing the mass of the car, it is possible to fully characterize its motion equation. Combining equations (2) and (4) gives $F = C \cdot m = 0.338\,N$ and $k = D \cdot m = 0.418\,N \left(\dfrac{m}{s}\right)^{-n}$. Finally, the car's motion equation can be written as follows

$$0.1736 \cdot a = 0.338 - 0.418 \cdot v^{0.74} \quad (5)$$

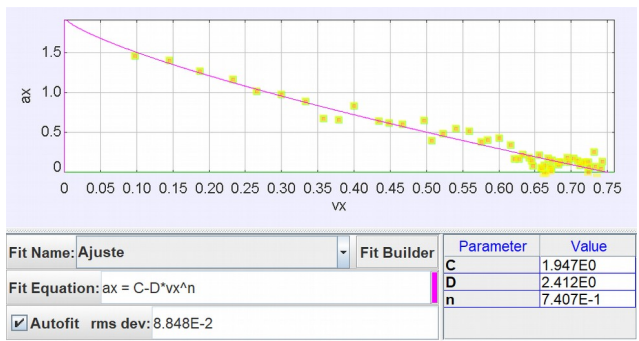

Figure 6. Acceleration as a function of the velocity obtained with *Tracker: points (experimental results) and non-linear fitting (red line)*.

### 4.2 Dynamical model

The *dynamical model* tool of Tracker was used to verify the predictive power of the model. Figure 7 shows a screenshot of the interface of the tool being used to create the model by entering the value of each of the parameters in the motion equation and the initial conditions for the car's motion. Detailed tutorials on how to create models can be found in the references [6, 13].

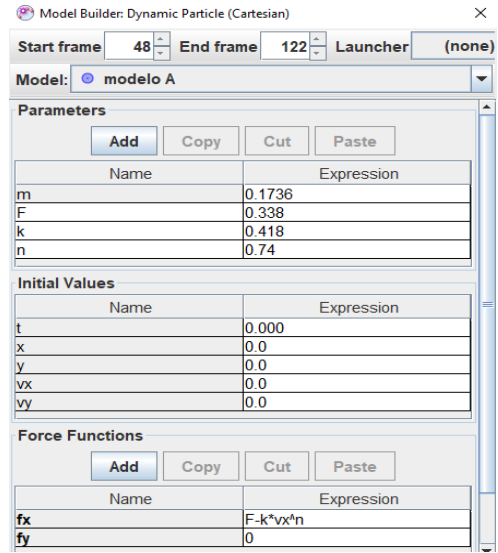

Figura 7. Screenshot of the *Tracker* Model builder.

Tracker numerically solves the differential equation of motion using the fourth-order Runge-Kutta method. The values obtained in this way were compared with those obtained with the *autotracker* tool. Figure 8 shows the position as a function of time (top) and the velocity as a function of time (bottom), with the experimental data shown in red and numerical results in blue. A high degree of concordance was found between the two data sets. This is hardly surprising in this case, in view of the closed-loop nature of the method—*i.e.*, the dynamical model was created by fitting the experimental *a(v)* data to a non-linear function.

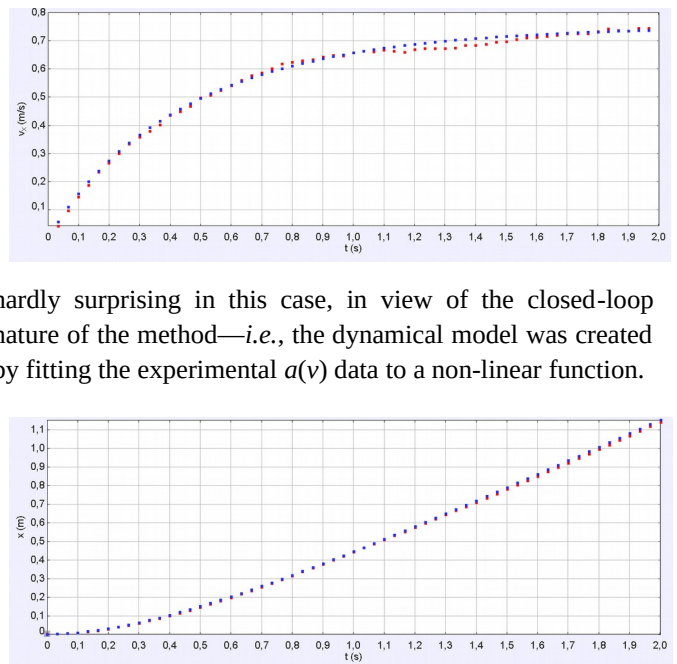

Figure 8. Experimental data (red) and numerical results (blue) for the position (top) and the velocity (bottom) as a function of time.

Using Tracker, it is possible to view a plot of numerical results in the same graphic interface as the experimental plots, allowing to compare the position measured experimentally with that obtained numerically at different times, as shown in Fig. 9.

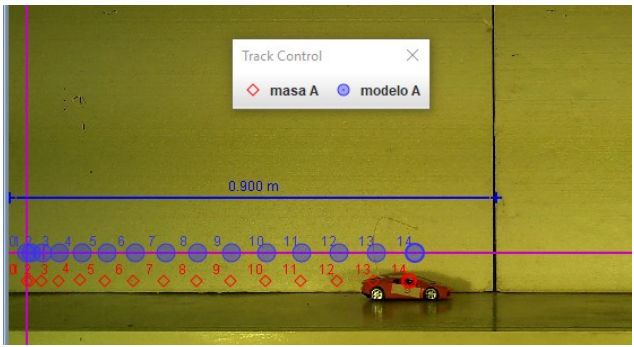

Figure 9. Tracker screenshot of the comparison between the experimental data and the numerical model.

## 5. Conclusions

The analysis of experimental systems whose differential equations of motion have no explicit analytical solution—an aspect seldom discussed in introductory courses—has great didactic value, for it is usually the case with real systems. In such situations, the formulation of models is a very powerful tool. It is worth mentioning that our numerical/practical approach provides an equation that works fine without needing to grasp in the details of the internal resistive forces involved.

Trajectories in the phase space are abstract constructs whose interpretation can prove conceptually very valuable as it allows students to qualitatively understand the temporal evolution of a system governed by a first-order differential equation, as would be the case of a falling object subjected to a velocity-dependent drag force or a variable-mass system.

Use of readily available computer tools like Tracker enables the analysis of experimental kinematic data in a phase space and the development of a dynamical model based on the numerical solution to the system's motion equation.

In a classroom setting, such tasks are found useful—as they aid in engaging students more actively—and encourage the learning process—as they allow students create their own models and verify their predictions through experiment. In addition, the experiment is inexpensive and can be carried out outdoors.

## Acknowledgements

We acknowledge financial support from grant FSED_3_2016_1_134232 (ANII-CFE). Translated from the Spanish by Eduardo Speranza.